\documentclass[%
 reprint,
 amsmath,amssymb,
 aps,
showkeys,
]{revtex4-2}

\usepackage{blindtext}
\usepackage{amsmath}
\usepackage{amssymb}
\usepackage{bbold}
\usepackage{verbatim}
\usepackage{booktabs}
\usepackage{graphicx}
\usepackage{dcolumn}
\usepackage{bm}

\begin{document}

\preprint{APS/123-QED}

\title{Machine Learning Without a Processor: Emergent Learning in a Nonlinear Electronic Metamaterial}

\author{Sam Dillavou}
    \email[Correspondence email address: ]{dillavou@sas.upenn.edu}
    \affiliation{Department of Physics and Astronomy, University of Pennsylvania, Philadelphia, PA, 19104, USA}

\author{Benjamin D. Beyer}
    \affiliation{Department of Physics and Astronomy, University of Pennsylvania, Philadelphia, PA, 19104, USA}

\author{Menachem Stern}
    \affiliation{Department of Physics and Astronomy, University of Pennsylvania, Philadelphia, PA, 19104, USA}

\author{Marc Z. Miskin}
    \affiliation{Department of Electrical and Systems Engineering, University of Pennsylvania, Philadelphia, PA, 19104, USA}

\author{Andrea J. Liu}
    \affiliation{Department of Physics and Astronomy, University of Pennsylvania, Philadelphia, PA, 19104, USA}
    \affiliation{Center for Computational Biology, Flatiron Institute, Simons Foundation, New York, NY 10010, USA}

\author{Douglas J. Durian}
    \affiliation{Department of Physics and Astronomy, University of Pennsylvania, Philadelphia, PA, 19104, USA}
    \affiliation{Center for Computational Biology, Flatiron Institute, Simons Foundation, New York, NY 10010, USA}

\date{\today}

\begin{abstract}
Standard deep learning algorithms require differentiating large nonlinear networks, a process that is slow and power-hungry. Electronic \textit{learning metamaterials} offer potentially fast, efficient, and fault-tolerant hardware for analog machine learning, but existing implementations are linear, severely limiting their capabilities. These systems differ significantly from artificial neural networks as well as the brain, so the feasibility and utility of incorporating nonlinear elements have not been explored. Here we introduce a \textit{nonlinear} learning metamaterial -- an analog electronic network made of self-adjusting nonlinear resistive elements based on transistors. 
We demonstrate that the system learns tasks unachievable in linear systems, including XOR and nonlinear regression, \textit{without a computer}. We find our nonlinear learning metamaterial reduces modes of training error in order (mean, slope, curvature), similar to \textit{spectral bias} in artificial neural networks. The circuitry is robust to damage, retrainable in seconds, and performs learned tasks in microseconds while dissipating only picojoules of energy across each transistor. This suggests enormous potential for fast, low-power computing in edge systems like sensors, robotic controllers, and medical devices, as well as manufacturability at scale for performing and studying emergent learning.
\end{abstract}


\maketitle

Arbitrarily complex functions can be produced by combining a sufficient number of even simple nonlinear operations~\cite{hornik_multilayer_1989}. Deep learning has exploited this fact, using digital computers to simulate ever larger artificial neural networks (ANNs) capable of elaborate and nuanced tasks in mechanical control, vision, and language~\cite{nguyen_review_2019, wang_yolov7_2022, baktash_gpt4_2023}. Training and performing these tasks is typically accomplished using sequential digital logic to perform billions or more nonlinear calculations. However, this process is slow and power-hungry compared to neural networks in the human brain~\cite{balasubramanian_brain_2021, Christensen_2022_2022}, where 100~billion neurons simultaneously use nonlinear physical processes as their base level of computation~\cite{sengupta_power_2014, jaeger_formal_2023}.

One aim of the field of \textit{neuromorphic computing}~\cite{mead_neuromorphic_1990} is to reduce the time and energy deficits of ANNs using hardware. A popular approach spreads computation across specialized devices and/or onto electrical or optical systems~\cite{kim_experimental_2015, hu_dotproduct_2016, yao_fully_2020, vadlamani_physics_2020, reuther_survey_2020, dutta_ising_2021, wang_implementing_2022, laydevant_training_2023, mei_inmemory_2023}, reducing time and/or power required to perform a trained task. However, this class of hardware is typically still trained using gradient descent (as are all-digital ANNs), a process that requires both digital processing and near-perfect modeling of any physical computation~\cite{wright_deep_2022}. In this regard the stacking of nonlinear functions is a disadvantage, as it can rapidly compound small errors in modeling. Another approach more explicitly imitates biological processes, \textit{e.g.} spiking neural networks, using electronic circuits~\cite{wang_supervised_2020, Christensen_2022_2022}. These networks are powerful at performing specific tasks but general-purpose architectures remain a challenge, while memory requirements for learning hinder scalability~\cite{Christensen_2022_2022}. 

A subfield of soft matter physics, \textit{physical learning}, aims to train physical systems to exhibit desired responses to stimuli~\cite{stern_learning_2023}. Unlike feed-forward ANNs, physical systems like mechanical or analog electrical networks can offer the advantage of being completely recurrent, with no imposed direction of information flow. However, in order for these systems to learn on their own, they must evolve during training by local rules~\cite{stern_learning_2023}. In some cases, natural physical dynamics can be hijacked to produced desired responses~\cite{pashine_directed_2019}, but understanding how complex responses can be trained remains an active problem~\cite{pashine_local_2021, kaspar_rise_2021, hopkins_using_2023}.

Recently, the fusion of neuromorphic computing with physical learning has led to a `physical learning machine' or `self-learning network' or `learning metamaterial,' a system capable of performing supervised learning tasks adaptively without any external processing~\cite{dillavou_demonstration_2022}. The system is a network of identical electronic components, namely variable resistors that self-adjust their resistances according to local rules, utilizing the Coupled Learning framework~\cite{stern_supervised_2021}, closely related to Equilibrium Propagation~\cite{scellier_equilibrium_2017} and Contrastive Hebbian Learning~\cite{movellan_contrastive_1991} (see \textit{Energy-Based Learning Frameworks} in the appendix for a comparison of these three frameworks). In such a system, supervised learning of nontrivial tasks is \textit{emergent}, in that it stems from the collective effects of many edges implementing local rules to self-adjust. Like hybrid digital-analog hardware~\cite{wright_deep_2022}, the ``forward computation" is performed by physical optimization, harnessing the principle of minimum power dissipation~\cite{vadlamani_physics_2020}; as voltage inputs are applied, natural physical dynamics lower the dissipated power, automatically ``calculating'' the voltage outputs. The learning metamaterial offers a distinct scalability advantage over hybrid digital-analog hardware because no companion simulation is required, so errors during training that arise from modeling the physical process do not compound with system size. Unfortunately, the variable resistor network suffers a fatal flaw -- it is linear. Linear networks like the original perceptron~\cite{Rosenblatt_1957} cannot perform simple nonlinear functions like XOR, regardless of how many elements they contain. 
In simulation, Equilibrium Propagation~\cite{scellier_equilibrium_2017} has been successfully leveraged to train nonlinear circuits meant to explicitly imitate and feed-forward ANNs by utilizing active elements~\cite{kendall_training_2020}, large voltage dynamic ranges~\cite{scellier_energybased_2023a}, and idealized circuit components ~\cite{scellier_energybased_2023a}. Further, these implementations rely on non-linearities (\textit{e.g.} diodes) that are distinct from than those presented in this work (transistors). Hybrid hardware-in-the-loop implementations are also emerging~\cite{drouhin_nanoelectronic_2023}, however they rely on centralized processing and use components not amenable to (CMOS) microfabriction. Thus, the feasibility and functionality of a physically implemented, potentially scalable, nonlinear learning metamaterial has not been demonstrated.

\begin{figure*}[t]
\centering
\includegraphics[width=2\columnwidth]{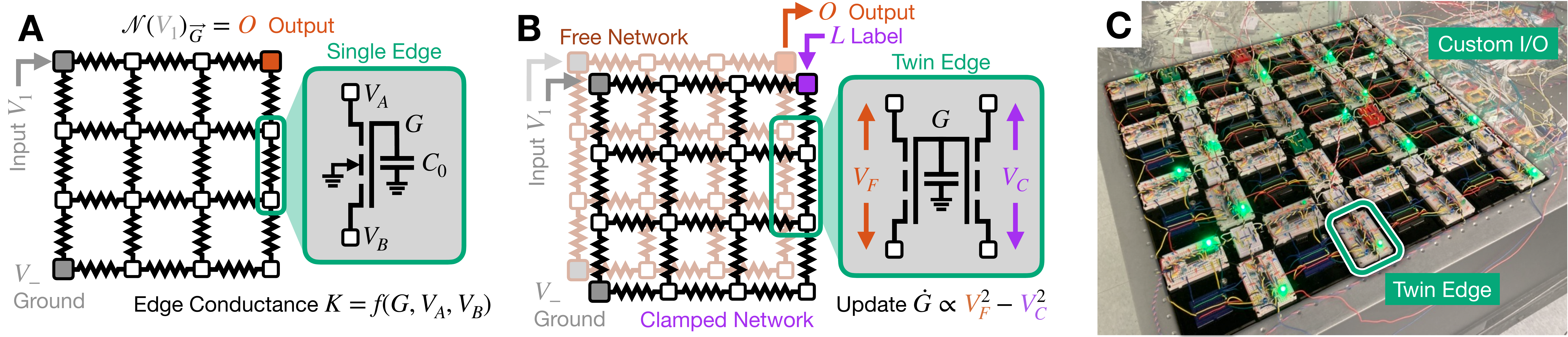}
\caption{\textbf{Design of the Learning Metamaterial} \textbf{(a) Physical response} We consider a physical system, an analog resistor network, as a function. Imposed voltages $V_1$ (input) and $V_-$ (ground) create a response throughout the network, and we choose a node to measure voltage output $O$. Our network is constructed using N-channel enhancement MOSFET transistors as variable nonlinear resistors. The conductance of each edge of the network thus depends on its connected node voltages $V_A$ and $V_B$, as well as the voltage on its gate $G$. $\vec G$ are the \textit{learning degrees of freedom} that change during the course of training, and are each stored on local capacitors with capacitance $C_0$. 
\textbf{(b) Learning Response} We duplicate this network, and overlay the copies such that they do not interact directly, but the transistors on commensurate edges draw their gate voltages $G$ from the same capacitor. This ensures two copies of the same electronic network. We designate one network as `Free' and impose only inputs ($V_1$, $V_-$), and the other as `Clamped' and impose inputs as well as the label (desired output) $L$. During training, circuitry on each twin edge continuously updates $G$ by charging or discharging the local capacitor, depending on the local difference between the electronic states in the two networks.
\textbf{(c) Image of the system} The highlighted `twin edge' is a transistor pair like the one shown schematically in (b), along with the circuitry to  update $G$. This twin edge is repeated 32 times to create our system. Nodes in the network contain no circuitry and are simply connections between twin edges, and the only interaction a computer has with the network is through the custom input/output (I/O) hardware (described in \textit{Supervisor and Measurement Circuitry} in the appendix, to impose boundary conditions ($V_1$, $V_-$, $L$), take measurements, and turn learning ($\dot G$) on and off. Note that the network has periodic connections, which are not pictured in the schematics in (a) and (b) for convenience. Custom I/O hardware, optical table, and all other aspects of the image except for the learning metamaterial itself are faded.}
\label{fig:0}
\end{figure*}

Here we report the laboratory implementation and study of a nonlinear learning metamaterial. The design may be microfabricated, establishing a paradigm for scalable learning that is distinct from that of biological brains. The system performs tasks using an electronic network of transistors wired as continuous and nonlinear variable resistors. During training the transistor gate voltages continuously evolve, encoding the learned state, until they are frozen by a supervisor. We detail the design of a single base element, which is repeated 32 times to create the autonomously trainable system. We demonstrate the system's ability to learn different nonlinear tasks, including regression and XOR, and show that it learns by fitting modes of the training data in order: mean, slope, curvature. Our system takes seconds to learn tasks, which it can then perform in microseconds while dissipating picojoules of power across each element.

\section*{System Description and Dynamics}

We construct an electronic network using N-channel enhancement MOSFET transistors as the edges, as shown schematically in Fig.~\ref{fig:0}A. We restrict the range of gate voltages $G$ such that these transistors operate in the \textit{Ohmic} or \textit{Triode} regime (see Twin Edge Circuitry in Appendix for details.) Here, the current $I$ passing between nodes may be modeled by the transistor `Square-Law,' as a sum of linear and quadratic terms of $\Delta V$ ($=V_A-V_B$ in Fig.~\ref{fig:0}). In this regime, the transistor may be treated as a passive nonlinear resistor, rewriting the Square-Law to give conductance $K\equiv I/\Delta V$
\begin{equation}
    K \propto G - V_{\textrm{th}} - \overline V
    \label{conductance}
\end{equation}
where $G$ is the gate voltage of the transistor, $V_{\textrm{th}} \approx 0.7$~V is a fixed threshold voltage, and $\overline{V} \equiv (V_A+V_B)/2$ is the average voltage at the two network nodes bookending the edge. Note that the dependence of conductance $K$ on $\overline{V}$ is the source of nonlinearity in our system. As a result, edges with lower gate voltages are both less conductive and more nonlinear (see \textit{Transistors as Nonlinear Resistors} in Appendix for details and a derivation of Eq.~\ref{conductance}).

Enforcing voltage boundary conditions to two or more nodes of the network creates a voltage state across the network. Designating one such enforced (`input') voltage as $V_1$ and the measured voltage at another node as the `output' voltage $O$, we can think of this network as evaluating a nonlinear function 
\begin{equation}
    O = \mathcal{N}(V_1)_{\vec G}
    \label{netout}
\end{equation}
where the form of $\mathcal{N}$ depends on network structure, any other boundary conditions (e.g. $V_-$ (ground) in Fig \ref{fig:0}a-b) and the gate voltages $\vec G$, one per edge, which serve as the \textit{learning degrees of freedom} in our system. These values are analogous to the \textit{weights} in machine learning: they evolve during training and encode the memory of that training. Eq.~\ref{netout}, \textit{inference} in machine learning, is performed by physics, meaning that successful training results in a passive network whose structure `solves' the task.

\begin{figure*}
\centering
\includegraphics[width=11.4cm]{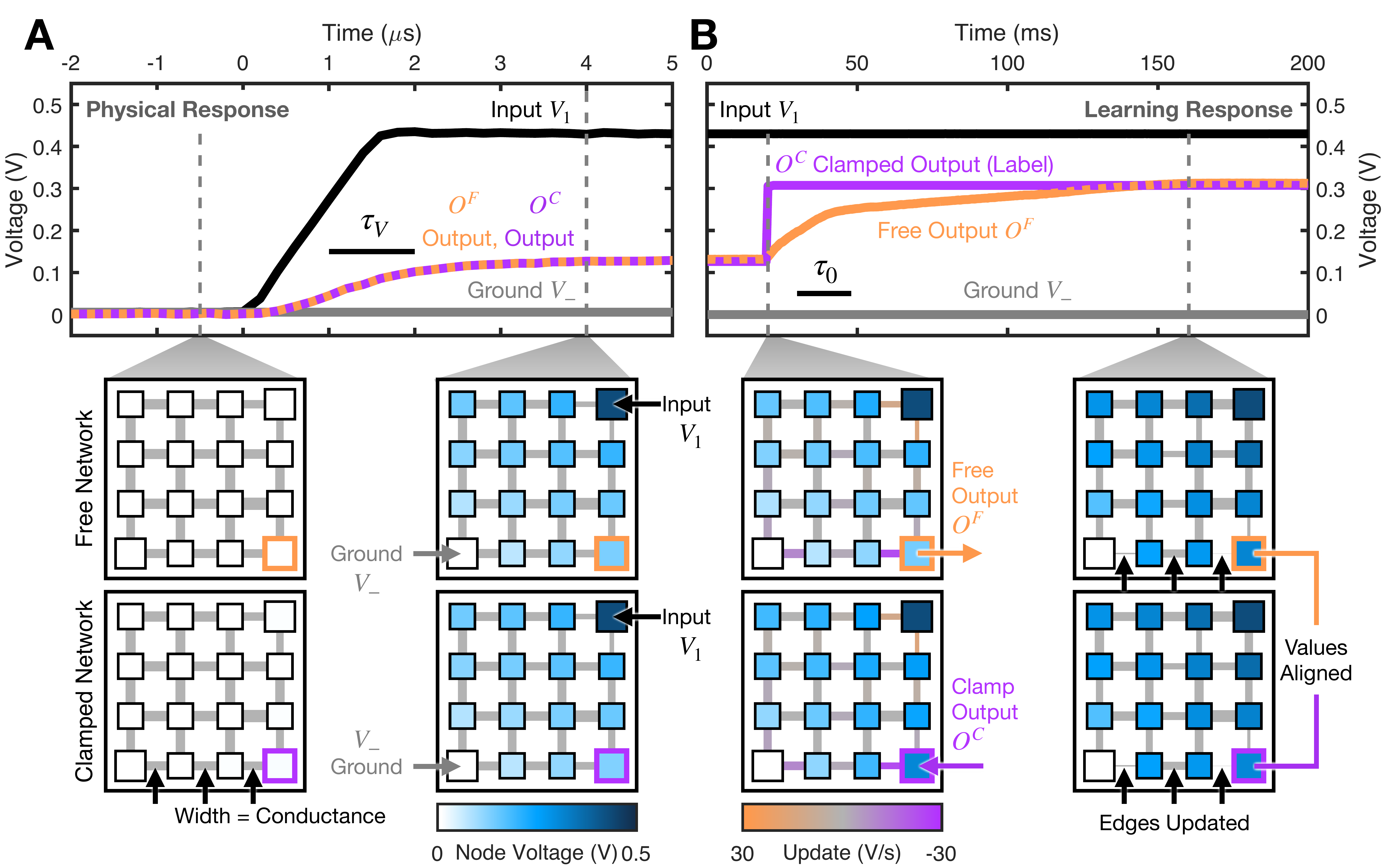}
\caption{\textbf{Emergent Supervised Learning} (a) Physical Response. When inputs are applied, nodes in the system find electronic equilibrium on a timescale of $\tau_V \sim 1~\mu$s. In this way, the network `calculates' the output (orange) naturally from the inputs. Note that the voltage change does affect the conductance of some edges (Eq.~\ref{conductance}), but not the learning degrees of freedom ($\vec G$) which are frozen. 
(b) Learning Response. Enforcing an output value in the clamped network (here $O^C=L=0.31$~V) and unfreezing the evolution of the gate voltages $\vec G$ will allow the system to continuously change these learning degrees of freedom. They will evolve with timescale $\tau_0 =18$~ms until frozen, or until the system reaches a state where the two networks have identical voltages, that is, the labels are naturally generated from the inputs, and the task has been learned. }
\label{fig:1}
\end{figure*}

We now follow the Coupled Learning framework~\cite{stern_supervised_2021} to construct dynamical rules that collectively solve the inverse problem $ L = \mathcal{N}(V_1)_{\vec G}$. In particular, the desired local update rule for an individual gate voltage $G$ is
\begin{equation}
 \dot G \propto \dot K \propto -\partial(P_C^\textrm{tot}-P_F^\textrm{tot})/\partial K = V_F^2-V_C^2
\end{equation} 
where $K$ is the conductance of that edge, $P^\textrm{tot}$ is the dissipated power for the entire system. For details, see \textit{Learning Rule Derivation} in the appendix. We note that this framework extends to functions with multiple inputs ($V_1$, $V_2$, etc) as well as multiple outputs \cite{dillavou_demonstration_2022}.  $V$ here is the voltage drop across an edge (\textit{i.e.} $V_A-V_B$), and subscripts $F$ and $C$ denote two electrical states of the same system, `free' and `clamped.'  To simultaneously access both states, we use a twin-network approach~\cite{dillavou_demonstration_2022, wycoff_desynchronous_2022, stern_physical_2022}, as shown in Fig.~\ref{fig:0}B. We construct two copies of the same transistor network, and pair them by connecting the gates of commensurate edges to a single capacitor, as shown in Fig.~\ref{fig:0}B. This ensures our copies will always evolve identically. During training, we apply the same inputs (here $V_1$, $V_-$) to both networks, and the label $L$ to the `clamped' network only.

Between every edge pair is a small circuit that runs current to the joint gate capacitor and changes its voltage according to
\begin{equation}
    \dot G =\frac{V_F^2-V_C^2}{V_0 R_0 C_0} 
    \label{Gdot0}
\end{equation}
where $V_F$ and $V_C$ are the voltage drops across the two transistors (`free' and `clamped'), as shown in orange and purple respectively in Fig.~\ref{fig:0}B, and $R_0 =100~\Omega$, $C_0 = 22~\mu$F, $V_0=0.33$~V are constants set by circuit components (see Twin Edge Circuitry in Appendix). We find it useful to extract a learning time constant $\tau_0 = R_0C_0(V_0G_{rng}/V_{max}^2)=18$~ms where $V_{max}=0.45$~V and $G_{rng}=5$~V are the approximate ranges of node and gate voltages permitted in the system, respectively.  Then the learning equation can be rewritten as
\begin{equation}
    \frac{\dot G}{G_{rng}} =\frac{V_F^2-V_C^2}{\tau_0 V_{max}^2} 
\label{Gdot}
\end{equation}
normalizing with node and gate voltage values. From Eq.~\ref{Gdot} it is clear that evolution becomes significant on the scale of $\tau_0$ and stops only when all free and clamped voltages are equal, meaning $O = L$ and the inverse problem is solved.

We typically \textit{nudge} the clamped output $O^C$ towards the desired label $L$ as instead of directly imposing it \cite{scellier_equilibrium_2017,stern_supervised_2021, dillavou_demonstration_2022}. The clamped output $O^C$ is then enforced as
\begin{equation}
    O^C = \eta L + (1-\eta) O
    \label{eta}
\end{equation}
where $0<\eta \le 1$ controls the size of the nudge. We may also define the difference between two nodes as our output $O = O_+-O_-$, as in \cite{kendall_training_2020}, which allows inhibitory input-output relationships. This scheme is clamped using
\begin{equation}
    O^C_\pm = O_\pm \pm \frac{\eta}{2} (L-O) 
    \label{diffclamp}
\end{equation}
which recreates Eq.~\ref{eta} when appropriately subtracted. Eq.~\ref{eta} (or \ref{diffclamp}) is implemented with a feedback circuit, described in \textit{Clamping Circuitry} in the appendix. As a result, $O^C$ evolves continuously with $O$. While infinitesimal $\eta \ll 1$ has typically been used in theory, in practice it creates an impractically small learning signal. We therefore use $\eta = 1$ (directly clamping the label) for all regression tasks (Fig.~\ref{fig:3} and \ref{fig:3B}), and $\eta = 0.5$ for Fig.~\ref{fig:4}, and find that $\eta \geq 0.2$ produces consistent results, as in past implementations~\cite{dillavou_demonstration_2022}.

Finally, we impose a global binary switch that toggles this evolution (Eq.~\ref{Gdot}) on and off. This is the only direct communication the supervisor has with the elements of the system. It allows us to freeze and measure the system and its functionality at any point, as well as to switch training data without unwanted effects. Freezing the system creates a passive, unchanging electrical network that can then be used for its trained purpose. For details about the circuitry used to implement learning and freezing, see \textit{Twin Edge Circuitry} in the appendix. By contrast with the original digital linear circuitry \cite{dillavou_demonstration_2022}, this new version is fully analog, operates without a clock, has nonlinear IV characteristics, and implements the exact Coupled Learning rule.

Once constructed, the learning metamaterial is a standalone system, and is trained purely by imposing boundary conditions in a two step process. First, a training example (datapoint) is chosen and applied to the network as boundary conditions - inputs to both networks and labels to clamped only - as shown in Fig.~\ref{fig:0}B. Next, learning is turned on for time $t_h$, and the gate voltages $\vec G$ evolve according to Eq.~\ref{Gdot}. This process is then repeated until training is complete, selecting a new datapoint each time. For all tasks in this work we use $t_h = 100~\mu$s, so that the system takes small steps in $\vec G$ space and can simultaneously learn many datapoints~\cite{dillavou_nonlinear_2023}.

Our system is a pair of self-adjusting twin networks, but it can also be viewed as an ensemble, composed of many copies of a single element, as shown in Fig.~\ref{fig:0}C. This element, the MOSFET transistor pair, self-adjusts according to its electrical state, and this local evolution across all 32 twin edges produces supervised learning as an emergent property. An example of this process for a simple (single datapoint) task is shown in Fig.~\ref{fig:1}. Input boundary conditions $V_-$ and $V_1$ are applied (0 and 0.43~V), generating equilibrium voltages across both networks, including output $O$ (Fig.~\ref{fig:1}A). A label $L=0.31$~V is imposed on the output node of the clamped network, and evolution is unfrozen, allowing edges to evolve under Eq.~\ref{Gdot}. The direction of evolution of each edge in this example is shown on the network schematic in Fig.~\ref{fig:1}B. The system evolves continually until learning is frozen or, as in this case, the the networks align, resulting in a steady state (Eq. \ref{Gdot0}). As the voltage equilibration timescale is much shorter than the learning timescale $\tau_V \approx 1 \mu s \ll \tau_0 = 18$~ms, this process works continuously, with voltages rapidly equilibrating when $\vec G$ are changed. In this and subsequent examples we report training time as the time spent unfrozen, ignoring time to freeze the system and to switch datapoints and equilibrate, if necessary. Including these times would approximately double training times, and this accounting is further described in the discussion section.

\section*{Results}
\begin{figure}[t!]
\centering
\includegraphics[width=8.7cm]{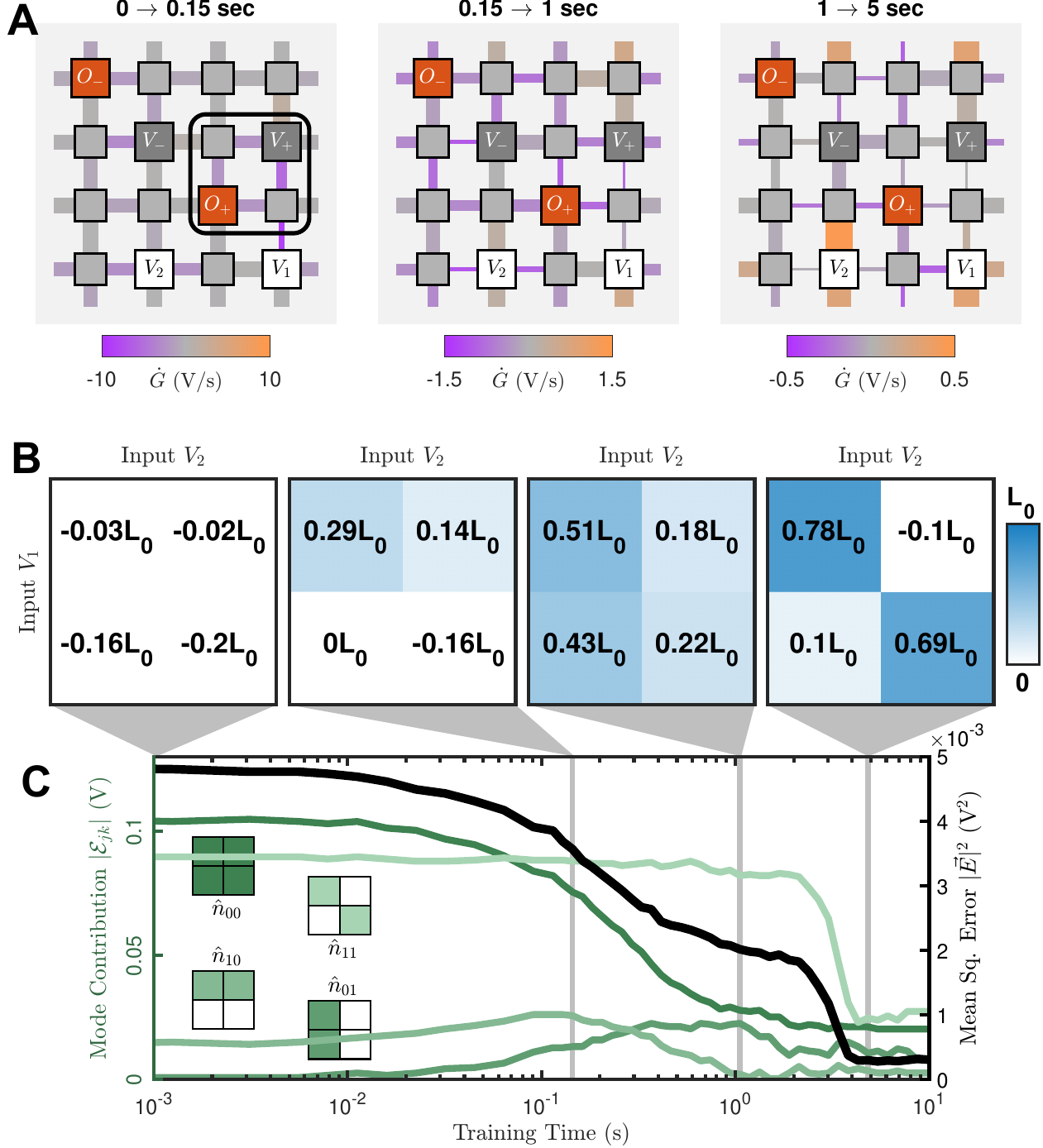}
\caption{
\textbf{Learning XOR} a) Schematic of the network showing input and output node locations and average rate of change in $G$ (edge color) for three time spans during training. Edge widths correspond to average conductance. The boxed region in the first panel highlights changes associated with altering the average output (reducing $|\mathcal{E}_{00}|$). b) Network output $O$ plotted as a function of inputs $V_1$ and $V_2$ in a truth-table format for four points during training. Color corresponds to output value, with $L_0 = -0.087$~V c) Mean squared error (black) and error contributions broken down by mode $|\mathcal{E}_{jk}|$ (green) over time. Error modes are depicted graphically next to lines. Time points indicated in (b) are denoted by vertical gray bars.
}
\label{fig:4}
\end{figure}

As the first demonstration with the above circuitry, we train our system to perform the canonical nonlinearly-separable function XOR. We choose the input schema shown in Fig.~\ref{fig:4}A, two variable inputs $V_1$ and $V_2$, and a differential output scheme $O = O_+-O_-$. We further impose constant voltages $V_-=0.11$~V and $V_+=0.33$~V on two nodes, which gives the system the ability to raise or lower the average output of the system, independent of the variable inputs. We find that a variety of constant values perform equally well. We train our system using four training datapoints (sets of boundary conditions), and a nudge parameter $\eta=0.5$. For reasons that will become apparent, it is useful to define the possible values of each input and label as vectors. Our training set consists of the values applied to each input node $\vec V_1 = [0,0,0.45,0.45]~V$ and $\vec V_2 = [0,0.45,0,0.45]$~V, and corresponding labels $\vec L = [0,L_0,L_0,0]$ where $L_0 = -0.087$~V. Note that these vectors each have four elements because we have four training datapoints, and that the desired output for datapoint $i$, $(\vec L)_i$, depends only on whether $(\vec V_1)_i \neq (\vec V_2)_i$, indicating that $\vec L$ cannot be written as a linear function of $\vec V_1$ and $\vec V_2$.

We train our system for ten seconds, randomly selecting training points at each step. The network first evolves rapidly, slowing over time until the network reaches its final state around five seconds, shown in Fig.~\ref{fig:4}A, right panel. During training, the network goes through a variety of output truth tables before arriving at the desired checkerboard pattern of XOR, as shown in Fig.~\ref{fig:4}B. We define a time-dependent vector of outputs $\vec O(t)$ for each point in the training set, and a vector of signed errors $\vec E(t) = \vec O(t)-\vec L$. The evolution of mean squared error $|\vec E(t)|^2$ to near zero, as shown by the black curve in Fig.~\ref{fig:4}C, demonstrates a successful learning process. 

We may further understand the learning dynamics by projecting the error $\vec E(t)$ into a new basis. We first create a series of vectors $\vec v_{jk}$, which are comprised of element-wise multiplication of chosen powers ($j$ and $k$) of each input. That is, $(\vec v _{jk})_i = ((\vec V_1)_i)^j   ((\vec V_2)_i)^k$. As examples, $\vec v_{00}$ is a vector in which every element is $1  =(\vec v _{00})_i=((\vec V_1)_i)^0   ((\vec V_2)_i)^0 $, and $\vec v_{10}$ is equal to $\vec V_1$, the values for input 1 in the training set: $(\vec v _{10})_i=((\vec V_1)_i)^1   ((\vec V_2)_i)^0 = (\vec V_1)_i $. We apply the Gram-Schmidt orthonormalization process to [$\vec v_{00}$, $\vec v_{01}$, $\vec v_{10}$, and $\vec v_{11}$] and create a complete orthonormal basis [$\hat n_{00}$, $\hat n_{01}$, $\hat n_{10}$, and $\hat n_{11}$]. Each  $\hat n_{jk}$ is summarized graphically in Fig.~\ref{fig:4}C and detailed in \textit{Orthonormal Basis Construction} in the appendix.

\begin{figure}[t!]
\centering
\includegraphics[width=8.7cm]{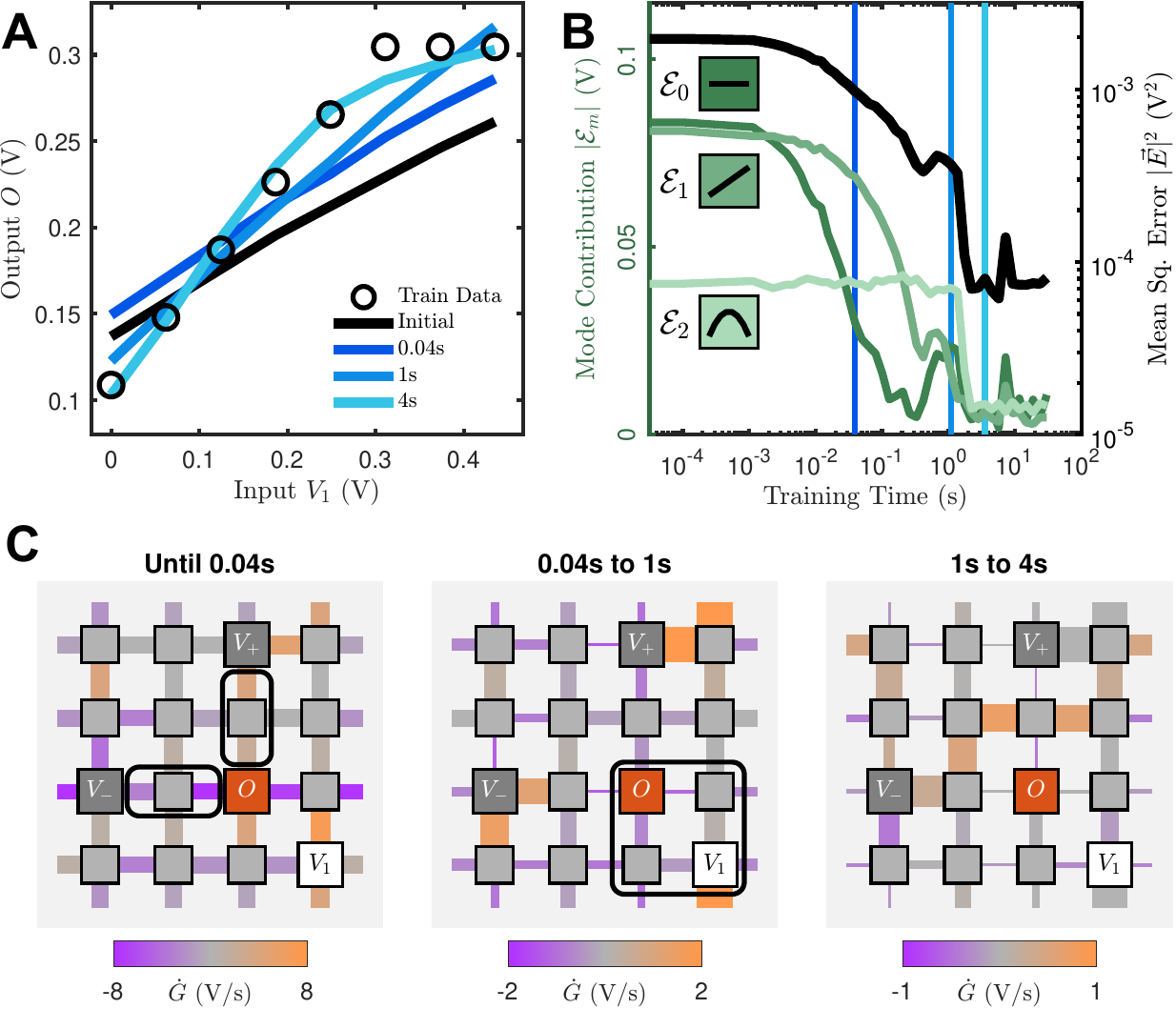}
\caption{\textbf{Nonlinear Regression}
a) Initial output (black line) and outputs at four points throughout training (blue lines) are shown versus variable input $V_1$ value. Training labels $L$ (black circles) are overlaid.
b) Mean squared error of all datapoints (black) and error contributions broken down by mode $|\mathcal{E}_m|$ (green) over time. Time points indicated in (a) are shown as vertical blue bars.
c) Schematic of the network showing input and output node locations and average rate of change in $G$ (color) for the three time spans during training defined by the blue lines in (a) and (b). Edge widths of the network correspond to average conductance of the edges. The boxed regions in the first two panels indicate interpretable changes discussed in the main text.
}
\label{fig:3}
\end{figure}

We calculate \textit{modes of error} $|\mathcal{E}_{jk}(t)| = |\hat n_{jk} \cdot \vec E (t) | $ over time, that is, the magnitude of the projection of error onto this new basis. We may interpret each mode $|\mathcal{E}_{jk}|$ as the portion of the total error that stems from specific contributions of each input: $\mathcal{E}_{00}$ represents the deviation from the incorrect average output, and $\mathcal{E}_{10}$ represents the error stemming from an incorrect linear relationship between $\vec V_1$ and $\vec O$. $\mathcal{E}_{11}$ represents the error from the (lack of a) dependence on the nonlinear combination of both input vectors, which is essential in XOR. We find that the system consistently adjusts these modes \textit{in order}, as shown in Fig.~\ref{fig:4}C, as well as for other tasks, as discussed below. In this task, we observe the change of $\mathcal{E}_{00}$, then $\mathcal{E}_{10}$ and $\mathcal{E}_{01}$. The lone nonlinear mode, $\mathcal{E}_{11}$, is the last to adjust, accompanying the final steep drop in error, and the desired checkerboard output table in Figs.~\ref{fig:3}B, right side.

Next, as a more complex task, we let our network train for nonlinear regression. We choose single nodes for our input $V_1$ and output $O$. To give the network access to the full range of output values regardless of input, we impose constant voltages on two nodes of the network $V_- = 0$~V and $V_+ = 0.45$~V. Our training dataset is eight elements, and $\vec V_1$ vs $\vec L$ is shown as black circles in Fig.~\ref{fig:3}A. When we initialize the network, $\vec O(t=0)$ is a line. During training, we cycle through datapoints in order (a random ordering gives identical results). In the first 4~s of training the network solves the task, as shown by sequential improvement of the the blue output curves in Fig.~\ref{fig:3}A. This progress follows a familiar pattern: first the mean is adjusted by shifting the line, then the slope is adjusted, and finally the line is bent. That is, the system learns the modes of the data in order, as it did for XOR.

To tease this out quantitatively, we again construct an orthonormal basis $\hat n_m$, this time with eight elements, like the training data. These are constructed by applying Gram-Schmit orthonormalization to $\vec v_m = (\vec V_1)^m$. We focus only on the first three vectors, $ \hat n_0$, $\hat n_1$, and $\hat n_2$, as we find higher modes negligible for this task, and further that our network is likely not deep enough to adjust them meaningfully. The construction of these vectors is detailed in the appendix under \textit{Orthonormal Basis Construction} in the appendix, but their shapes can be simply summarized as flat, linear, and parabolic when plotted against $\vec V_1$.

As shown in Fig.~\ref{fig:3}B, error in this basis $|\mathcal{E}_m|$ decreases in mode order: constant $(\mathcal{E}_0)$, then linear $(\mathcal{E}_1)$, then parabolic $(\mathcal{E}_2)$. These changes can be visually confirmed by noting the output shapes in Fig.~\ref{fig:3}A, whose times are plotted as vertical blue bars in Fig.~\ref{fig:3}B. After these modes decrease, the mean squared error $|\vec E(t)|^2$ plateaus. The constant and linear modes may further be identified with behavior of individual edges of the system during training, highlighted by the boxed regions in Fig.~\ref{fig:3}C.  In the first 0.04~s of training time $\mathcal{E}_0$ is eliminated by modifying the connections between the output and the constant inputs. From 0.04 to 0.4~s, the network reduces $\mathcal{E}_1$, by increasing the connection between the output and $V_1$, relative to its connections to the constant inputs (which both decrease).

For this and all tasks shown in this work, the network is overparameterized, and therefore has access to many directions in $\vec G$ space that do not affect the output. In practice, this means that the system is pushed along these dimensions by \textit{any} imperfection in the update circuitry, no matter how small. We find this bias to vary between edges and for individual edges between tasks, belying a voltage-state dependence. At long times ($>1$~s), the result is often many extreme gate voltage values (high and low) as shown in Fig.~\ref{fig:3}C, and difficulty interpreting specific changes to the network. While the network can and does find solutions in spite of these imperfections, bias sets an inevitable error floor. It is important to note however, that because each edge evolves according to its own local update rules, these errors do not compound. That is, unlike differentiating an entire simulation of a physical system, they do not require tighter and tighter precision as the system scale increases.

\begin{figure}[t]
\centering
\includegraphics[width=8.7cm]{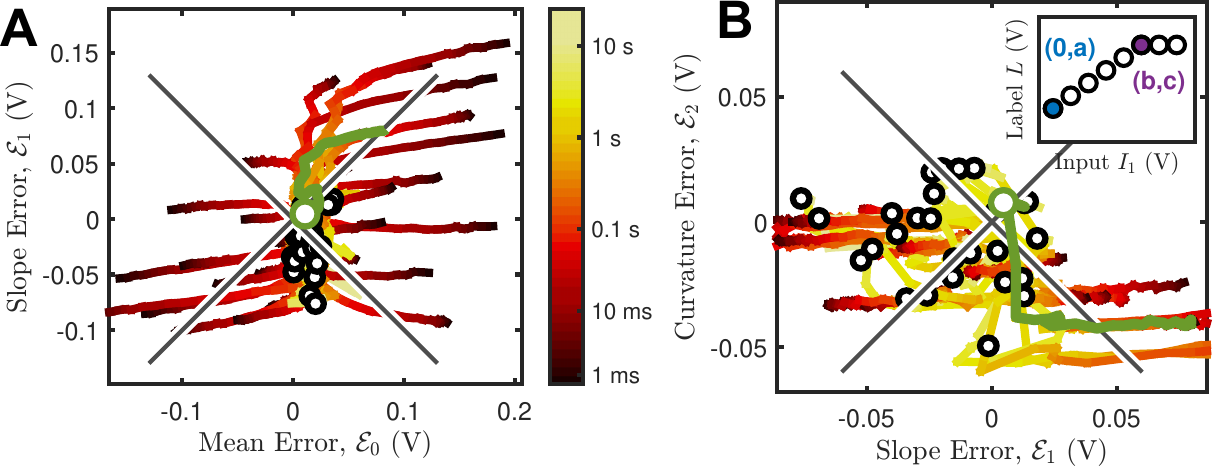}
\caption{\textbf{Error Mode Evolution for Many Tasks}
a) Signed error contributions from the lowest two modes $\mathcal{E}_0$ and $\mathcal{E}_1$ for 29 unique regression tasks during training. Input values and network setup are identical to Fig.~\ref{fig:3}. Color indicates training time, and traces end at hollow black circles. Gray diagonal lines represent $|\mathcal{E}_0| = |\mathcal{E}_1|$. Green line corresponds to the experiment shown in Fig.~\ref{fig:3}.
b) Same as (a) but for linear $\mathcal{E}_1$ and curvature $\mathcal{E}_2$ modes. 
Inset: training data for each of these 29 tasks is 8 datapoints in a piece-wise linear form like the one shown here. Parameters $a,b,c$ are varied (see text) to produce a range of initial error modes.
}
\label{fig:3B}
\end{figure}

We find this sequential mode reduction picture to be consistent across tasks. To demonstrate, we compare the error mode contributions to each other over time for 29 different nonlinear regression tasks, as shown in Fig~\ref{fig:3B}A ($\mathcal{E}_0$ vs $\mathcal{E}_1$) and \ref{fig:3B}B ($\mathcal{E}_1$ vs $\mathcal{E}_2$), with the same initialization of the circuitry. Each task is piece-wise linear with 8 datapoints and the same node setup and initial conditions as Fig.~\ref{fig:3}A (the data from Fig.~\ref{fig:3}A is highlighted in green). The tasks all share the common features of a positive linear slope from $(0,a)$ to $(b,c)$, and a flat line afterwards, as detailed in \ref{fig:3B}B inset. The tasks included each have a unique parameter combination, within the ranges of $a$ = [0.07, 0.24]~V, $c$ = [0.13, 0.30]~V, and with $b$ taking one of two values 0.12~V or 0.31~V. These parameters were chosen to vary the initial mode error contributions, while keeping $\mathcal{E}_0<0.2$~V to allow entire trajectories to be shown on a single plot. Note that $c>a$ for all tasks to ensure a positive slope. 

When nearly any zero mode $\mathcal{E}_0$ is present, its magnitude is decreased first, as shown by the shape of the trajectories in Fig.~\ref{fig:3B}A. The linear error mode $\mathcal{E}_1$ then shrinks, followed by the curvature contribution $\mathcal{E}_2$, as shown in Fig.~\ref{fig:3B}B. In this single-output configuration, the network is limited in its functionality (no inhibitory connections using $O_-$), and thus tasks the network can accomplish have small $\mathcal{E}_2$ contributions. Note that nonzero final error modes are at least in part caused by the aforementioned biases in each edge's learning rule evaluation.

\section*{Discussion}

To our knowledge, our learning metamaterial is the only physical implementation of a standalone system that can adaptively perform nonlinear supervised learning without a processor. Unlike previous implementations~\cite{dillavou_demonstration_2022, wycoff_desynchronous_2022, stern_physical_2022}, this circuit solves nonlinear tasks, varies its learning degrees of freedom continuously, and implements the full (not approximate) Coupled Learning rule. Our system learns by tackling modes of the data in order, much as if it is peeling layers of an onion, similar to the \textit{spectral bias} of ANNs which learn simpler functions faster than more complex ones~\cite{rahaman_spectral_2019a}. In the infinitesimal $\eta \ll 1$ limit, Coupled Learning aligns well~\cite{stern_supervised_2021} with gradient descent, the algorithm used in much of deep learning. However, it is interesting that the system learns in mode order even for the $\eta=0.5$ or 1 used in this work. 

Our circuitry learns tasks adaptively in seconds, and performs them power-efficiently. Training times reported in this work are integrated time spent learning (unfrozen). An approximately equal amount of time is spent applying voltages, but this can be easily reduced to a negligible factor with modestly better control hardware (\textit{e.g.} 1~MHz versus the current $\approx$100~kHz, see \textit{Supervisor and Measurement Circuitry} in appendix). Future versions may also realize large speed gains by shortening adjustable timescales of the system ($t_h = 100~\mu$s, $\tau_0=18$~ms). The circuitry used for learning is power hungry ($\sim$100~mW/edge) but may be completely shut off for inference. Practically, the storage of analog gate voltages provides poor memory fidelity over time, and future versions will likely require on-edge digital storage of these values, as in previous implementations~\cite{dillavou_demonstration_2022} to realize efficient use. Given these caveats, when performing the trained tasks in this work, our system dissipates approximately 10-20~pJ across each edge, a number calculated solely from measured voltage and conductance values and the equilibration time $\tau_V$, a useful lower bound for our prototype design as-is. We note however that unlike digital systems, a 100x speed-up ($\tau_V \rightarrow 10$ ns), plausible with the greatly decreased capacitance of microfabricated circuitry, would confers a 100x decrease in energy use, resulting in order 100~fJ per edge, equivalent to state-of-the-art analog AI accelerators~\cite{ambrogio_analogai_2023}. For details on these calculations see \textit{Power Dissipation} in the appendix, as well as~\cite{dillavou_nonlinear_2023} for performance on tasks with more datapoints). 

Our circuitry's design has several features that make scalability a realistic possibility. First, because physics performs the forward computation (Eq.~\ref{netout}), and updates (evolution) are calculated by each element in parallel (Eq.~\ref{Gdot}), the duration of a training step is independent of system size, provided $\tau_V \ll t_h$. By contrast, for ANNs, including neuromorphic implementations~\cite{wang_implementing_2022, yao_fully_2020, hu_dotproduct_2016, reuther_survey_2020, kim_experimental_2015, laydevant_training_2023, mei_inmemory_2023} and hybrid analog-digital pairings~\cite{hopkins_using_2023, wright_deep_2022}, memory requirements and training step duration scale with system size. Second, our design uses standard circuitry components amenable to microfabrication, allowing large system sizes ($\sim 10^7$ edges in $\sim 10$~mm$^2$ with well-established 180~nm CMOS technology~\cite{dillavou_demonstration_2022}) and reduction of equilibration timescale $\tau_V$. Third, unlike centralized computing devices, our system would not require individually-addressable components, and supervising the system does not get more complex with scale. Finally, it is robust to damage~\cite{dillavou_demonstration_2022}, making it is also fault tolerant in the manufacturing process. By contrast, manufacturing error is a major impediment to scaling up top-down processor-based chips for other neuromorphic hardware realizations~\cite{Christensen_2022_2022}.

Because of its scalability, low power, and speed, future versions of nonlinear learning metamaterials may compete with artificial neural networks for some applications. However, many interesting issues need to be addressed to reach that stage. For example, while the nonlinear ReLU activation function is known to suffice for ANNs, it remains unclear what kind of nonlinearity in the IV characteristics of the resistive edges is needed in general, much less what kind is optimal, for accomplishing nonlinear tasks with Coupled Learning. Moreover, for these fully recurrent networks, the optimal network topology for a broad range of supervised tasks remains to be determined. The layered structure so important for backpropagation in feed-forward networks, often with high connectivity between layers, may not be needed. On the other hand, the square lattice used in our system is likely too simple and sparse. A hierarchical connectivity with long-range connections may be needed to efficiently communicate analog signals and avoid poor local minima, evocative of skip connections in ANNs and neural structure in the brain. The efficacy of more complex architectures, as well as how training time and energy scale with system size, remain important unanswered questions.

From a fundamental point of view, our self-learning system provides a unique opportunity for studying emergent learning. In comparison to biological systems including the brain, our system relies on simpler, well-understood dynamics, is precisely trainable, and uses simple modular components. Our current system has already been used to characterize a competition between energy use and precision~\cite{stern_training_2024}, a biologically-relevant trade-off, used for example by the neocortex during starvation in mice~\cite{padamsey_neocortex_2022}. Physical self-learning circuits may also be more explainable than ANNs. Their properties are sculpted by the interaction of the training process with physical constraints (\textit{e.g.} power minimization), leaving physical signatures of learning~\cite{rocks_revealing_2020, stern_physical_2024} that give valuable microscopic insight into emergent learning.

While the collective behavior of many-body systems differs fundamentally from the behavior of one body, the behavior of \textit{very many} bodies is generally similar to that of \textit{many} bodies. This is what makes simulation of relatively small systems useful for understanding condensed matter. Brains are a form of matter where the behavior of \textit{very many} is extremely different from that of \textit{many}. This can be seen simply by comparing the cognitive abilities of \textit{C. elegans} (302 neurons) to those of humans ($10^{11}$). \textit{Very many} differs from \textit{many} because the number of learning degrees of freedom (\textit{e.g.} synaptic strengths) increases with the number of bodies (neurons). Our system shares this property: the number of learning degrees of freedom (gate voltages) scales with the number of bodies (network edges). The speed and scalability of learning metamaterial systems imply that we may someday be able to study emergent physical learning on scales from \textit{many} to \textit{very many}, a feat that is impossible \textit{in silico}.

\acknowledgements 
    S.D. would like to thank Kieran Murphy for helpful discussions. This research was supported by the National Science Foundation via the UPenn MRSEC/DMR-1720530 and MRSEC/DMR-DMR-2309043 (S.D. and D.J.D.), and DMR-2005749 (A.J.L.), the Simons Foundation \# 327939 (A.J.L.), and the U.S. Department of Energy, Office of Basic Energy Sciences, Division of Materials Sciences and Engineering award DE-SC0020963 (M.S.). S.D. acknowledges support from the University of Pennsylvania School of Arts and Sciences Data Driven Discovery Initiative. M.Z.M. acknowledges support from the Packard Foundation. D.J.D. and A.J.L. thank CCB at the Flatiron Institute, as well as the Isaac Newton Institute for Mathematical Sciences under the program ``New Statistical Physics in Living Matter" (EPSRC grant EP/R014601/1), for support and hospitality while a portion of this research was carried out.

\bibliography{bib,extras}

 \appendix

\begin{figure}[b]
\centering
\includegraphics[width=8.7cm]{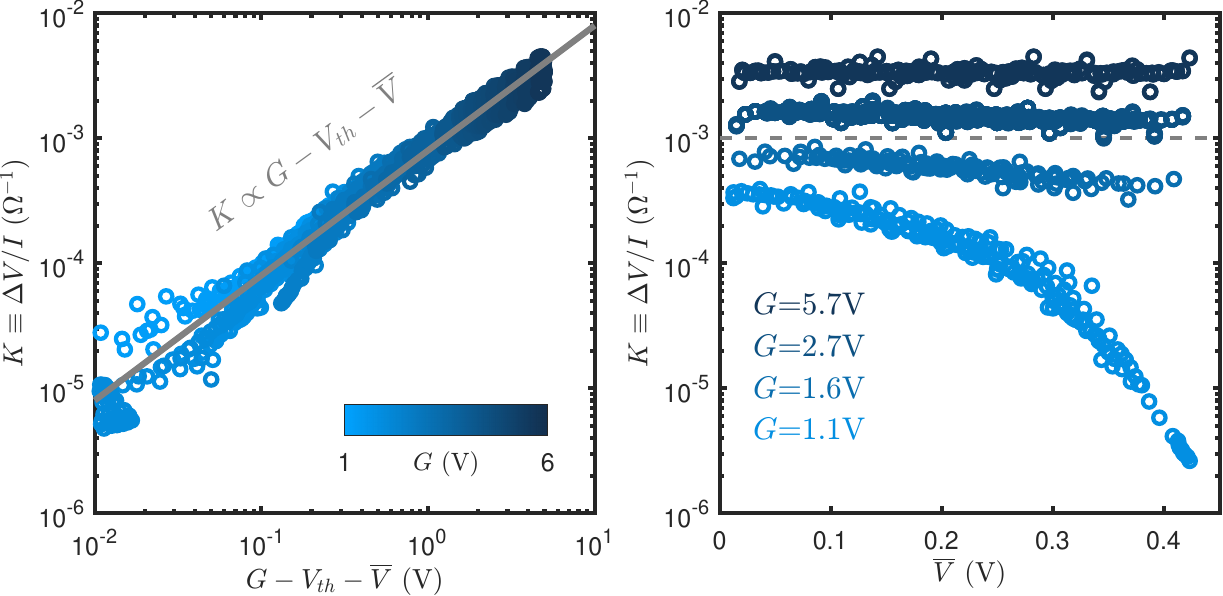}
\caption{\textbf{Transistors as Nonlinear Resistors} Left: Transistor source-drain conductance $K$ as a function of gate voltage $G$, threshold voltage $V_{\textrm{th}}\approx 0.7$~V, and average of source and drain voltages $\overline V$. Color is gate voltage, and the gray line is a guide for the eye of constant proportionality. Right: Transistor source-drain conductance as a function of average of source and drain voltages. Color represents gate voltage, with the same scale as left side. The dashed line represents a 1~k$\Omega$ linear resistor, which has the same conductance regardless of the voltages it experiences.}
\label{fig:nonlinear}
\end{figure}

\begin{figure*}
\centering
\includegraphics[width=17.8cm]{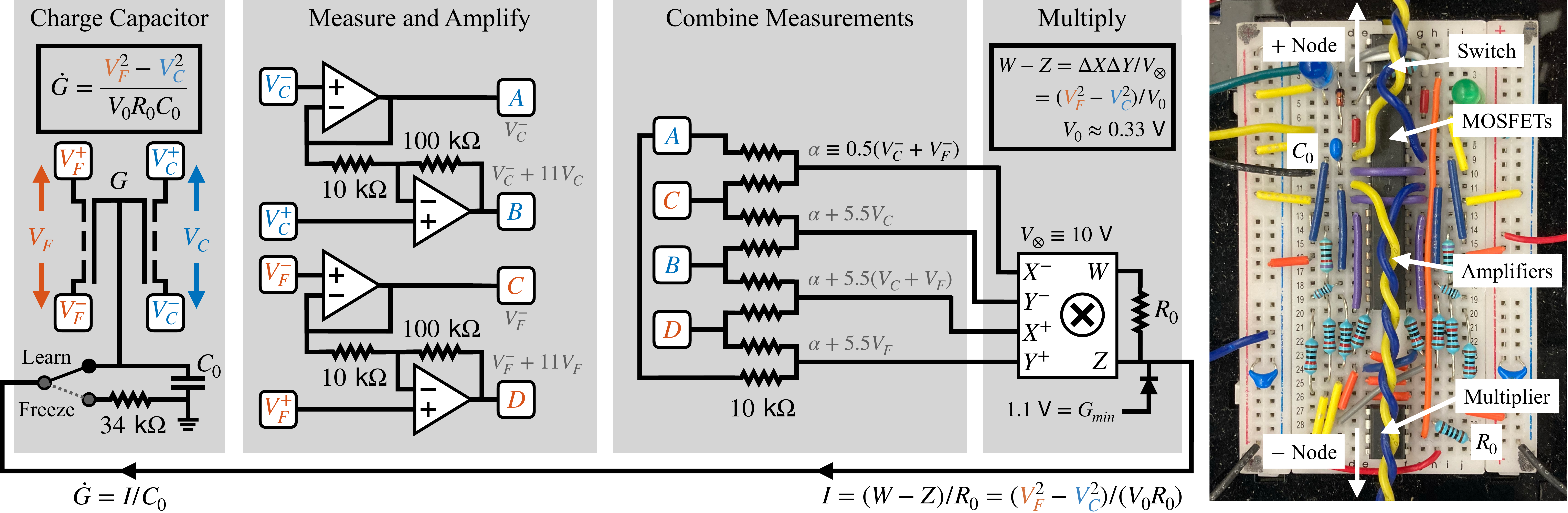}
\caption{\textbf{Twin Edge Circuitry} Schematic and image of the circuitry that implements the learning rule, which is the entirety of the operational elements of each twin edge (informational LEDs and power supplies are excluded). Starting from the left, The twin edge is comprised of two N-channel enhancement MOSFETS (ALD1106PBL) with body tied to ground (not shown). The output of the learning circuitry is a current which is fed into a charging capacitor $C_0 = 22~\mu$F, that holds gate voltage $G$. A switch (MAX322CPA+) can freeze the learning by instead shunting this current to ground. To start the calculation process, the voltages at the adjacent nodes in both networks are measured (one is arbitarily designated $+$ and the other $-$). The free and clamped measurements are combined and amplified using operational amplifiers (TLV274IN), then averaged in pairs and multiplied together using an analog multiplier (AD633ANZ). Note that $V_\otimes \equiv 10$~V is a property of the multiplier circuitry, and $V_0 \approx 0.33$~V is an amalgamation of factors from all stages of the circuitry (see Eq.~\ref{V0}). The multiplier output is wired to produce a current $(W-Z)/R_0$, completing the loop. $R_0 = 100~\Omega$.  Note that gate voltages are prevented from decreasing indefinitely using $G_{\textrm{min}}=1.1$~V and a diode (1N4148). The image on the right shows a physical edge, with each integrated circuit, charging capacitor, charging resistor, and adjacent nodes labeled.}
\label{fig:edgecircuitry}
\end{figure*}

\section{Transistors as Nonlinear Resistors}
In the Ohmic or Triode (passive) regime, the current between the source and drain $I$ in our N-channel enhancement MOSFETs (ALD1106PBL) obeys the transistor `Square-Law':
\begin{equation}
    I \propto \left[G-V_{th}-V_D \right ] V_{DS} + \frac{V_{DS}^2}{2}
    \label{squarelaw}
\end{equation}
where $G$ is the gate voltage, $V_{\textrm{th}} \approx 0.7$~V is the threshold voltage inherent to the transistors, $V_D$ is the voltage at the drain and $V_{DS}$ is the voltage drop from drain to source. We define $\overline V$ as the average voltage of the source and drain, or equivalently $V_A$ and $V_B$ in Fig.~1A, or $V^+$ and $V^-$ in Fig.~6, left side. Because our MOSFETs' bodies are tied to ground, there is no \textit{a priori} determination of source and drain, but rather the drain is the terminal with the higher voltage, \textit{i.e.} $V_D = \max(V_A,V_B)$,  $V_S = \min(V_A,V_B)$, allowing our edges to be bidirectional. We define the conductance across the source-drain $K \equiv \Delta V / I$, as voltage drop divided by current. We note that $V_{DS}/2-V_D = -(V_D+V_S)/2 = -\overline V$, and that $V_{DS} = \Delta V$ as defined in the text, allowing us to rewrite Eq.~\ref{squarelaw} as
\begin{equation}
\begin{split}
    I  \propto \left[G-V_{th}-\overline V \right ] \Delta V \\ \rightarrow K \equiv I/\Delta V \propto G-V_{th}-\overline V 
\end{split} 
\end{equation}
recovering Eq~\ref{conductance}.

Experimentally, transistor conductance is measured by imposing a gate voltage, as well as a voltage drop across the transistor source-drain in series with a known linear resistor, ranging from $R= 100~\Omega$ to $100~\textrm{M}\Omega$. Current across the transistor is calculated as the voltage drop across the resistor divided by its resistance. Results of these tests demonstrate a clear proportionality, as shown in Fig.~\ref{fig:nonlinear}, left. They further demonstrate nonlinearity, which comes in the form of $\overline V$ dependence, shown in Fig.~\ref{fig:nonlinear}, right. In our system, nonlinearity is coupled with conductance, and thus the network must push relevant edges into a low-conductance regime to solve nonlinear tasks.

\section{Twin Edge Circuitry}
The circuitry for a twin edge (the fundamental unit of our system) is described and shown schematically and in an image in Fig.~\ref{fig:edgecircuitry}. The learning timescale $\tau_0 \approx 18$~ms is a function of circuit components and of the voltage ranges allowed for nodes $V_{max} = 0.45$~V and gates $G_{rng}=5$~V, 
\begin{equation}
    \tau_0 \equiv \frac{G_{rng}}{V_{max}^2}V_0 R_0 C_0  \approx 18 \text{ ms}
\end{equation}
where $R_0 =100~\Omega$, $C_0 = 22~\mu F$, and $V_0$ is set by the gain in the node measurement amplification, a factor of 2 for averaging, and the analog multiplier (see Fig.~\ref{fig:edgecircuitry}):
\begin{equation}
    V_0 =  \left (\frac{10~\textrm{k}\Omega }{10~\textrm{k}\Omega + 100~\textrm{k}\Omega} \times 2 \right ) ^2\times V_\otimes \approx 0.33~\textrm{V}
    \label{V0}
\end{equation}
Where $V_\otimes = 10$ is a property of the analog multiplier.

\section{Orthonormal Basis Construction}

\begin{figure*}
\centering
\includegraphics[width=11.4cm]{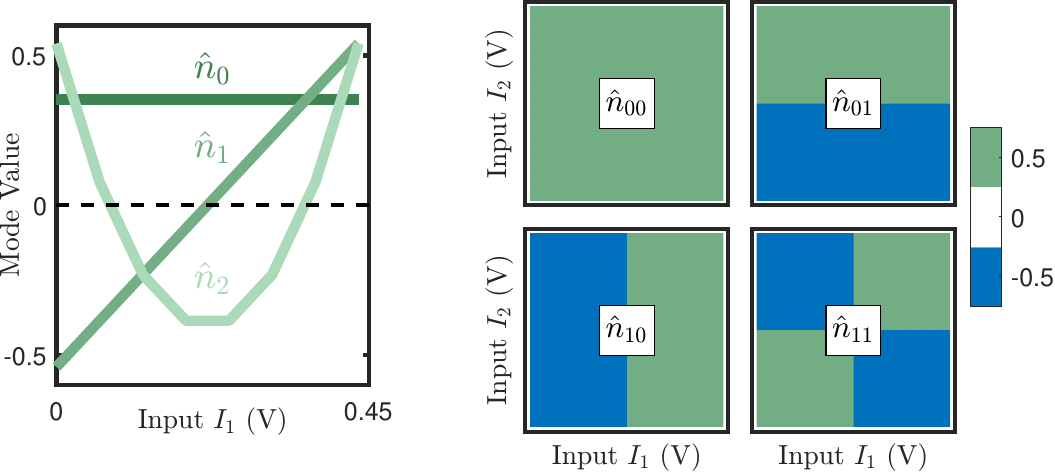}
\caption{\textbf{Orthonormal Modes} Left: Visualization of the modes for one variable input with eight equally-spaced input values between 0~V and 0.45~V (Fig.~4 and 5). Right: Visualization for two variable inputs, each with value 0~V or 0.45~V  (Fig.~3). Construction is described in the text.}
\label{fig:modes}
\end{figure*}

The orthonormal basis $\hat n_m$ used to analyze how our system learns nonlinear regression in Fig.~4 and 5 is constructed in order as follows. We take $\vec V_1$ to be the vector of all values applied at the variable input node in the training set, which for these two figures is 8 evenly spaced values ranging from $0$ to $V_{max}=0.45$~V. Let $\vec V^k_1$ be equal to the same vector with each element raised to the $k^{th}$ power. Then, for each mode $m$ in order, we apply the Gram-Schmidt process to create an orthonormal basis $\hat n_m$:
\begin{equation}
    \vec n_m = \vec V^m_1 - \left ( \sum_{k=0}^{m-1} (\vec V^k_1 \cdot \hat n_k) \hat n_k \right )
\end{equation}
\begin{equation}
    \hat n_m = \vec n_m / |\vec n_m|
\end{equation}
These basis vectors are shown plotted against $\vec V_1$ in Fig.~\ref{fig:modes} in green.

For the two-input case of XOR, we define $\vec V^{jk}_{12}$ the vector whose $i^{th}$ element is the $i^{th}$ element of $\vec V^j_1$ and of $\vec V^k_2$ multiplied together. We then perform Gram-Schmidt as before:

\begin{equation}
    \vec n_{jk} = \vec V^{jk}_{12} - \left ( \sum_{J=0}^{k-1}\sum_{K=0}^{k-1} (\vec V^{JK}_{12} \cdot \hat n_{JK}) \hat n_{JK} \right )
\end{equation}
\begin{equation}
    \hat n_{jk} = \vec n_{jk} / |\vec n_{jk}|
\end{equation}
Which subtracts away any parallel components to lower modes and normalizes the vectors. The modes used in Fig.~3 are plotted as heatmaps in Fig.~\ref{fig:modes}. Because only four datapoints are used in this case, four modes (00, 01, 10, 11) form a complete basis.

\section{Power Dissipation}
\subsection{Inference}
Here we calculate the power dissipated during the performance of a trained task (inference).
For the XOR task in Fig.~3, the final equilibrium state power dissipation, calculated as $K \Delta V^2$ across each edge, is 151~$\mu$W. For the 29 tasks in Fig.~5, the value is 345$\pm$72~$\mu$W. These equate to approximately 5 or 10~$\mu$W per edge, respectively. Note that we have not included other factors such as the power required to run the voltage application hardware, or the capacitance of breadboards and wires, but this is still a useful lower bound. Voltages in our circuit equilibrate in approximately 4~$\mu$S or less, certainly slowed by our voltage application hardware (see Fig.~2). As dissipation is low at the start of equilibration, we estimate energy cost by assuming 2~$\mu$S of equilibrium-level power dissipation. This leads to 10 or 20~pJ of dissipation per edge for these tasks, respectively.

As a reference point, we calculate the power-efficiency of running a standard feed-forward neural network on the most efficient supercomputer on the Green500 list\footnote{https://www.top500.org/lists/green500/2023/06/}. Using the quoted value $\sim$65 GigaFLOP/s/Watt = 65 GigaFLOP/Joule we find 15~pJ per FLOP (\textbf{F}loating \textbf{P}oint \textbf{OP}eration). Feed-forward neural networks perform inference using of order 1 FLOP/parameter, resulting in approximately 15~pJ per parameter. Note that digital supercomputers use at least 32 bits for each parameter (to achieve floating point accuracy), which in our system would equate to gate voltages accurate to single nano-Volts, well beyond our circuitry's reach. However, our system's ability to fine-tune gate voltages is at the very least superior to our 8 bit measurements; we often see improvements in final output as a result of collective changes in $\vec{G}$ below our measurement resolution.

A sharper comparison for our system is Qualcomm's state-of-the-art AI chips\footnote{https://www.qualcomm.com/content/dam/qcomm-martech/dm-assets/documents/Cloud AI 100 MLPerf 2.0 inference performance 2022-05-16.pdf}, which use approximately 1~pJ per \textbf{M}ultiply \textbf{AC}cumulate (MAC) operation. This figure can again be equated to approximately 1~pJ per parameter in inference in a system that uses 8 bits. This is within striking distance of our $\sim$15~pJ per parameter achieved with solderless breadboards and off-the-shelf parts. Further, our system can be made 100x more efficient with a 100x increase in speed, unlike digital systems which cost energy per operation. When micro-fabricated using state-of-the-art components with a far faster control circuit, such gains may be realized even with our current network design, allowing energy use of  150~fJ per parameter, approximately equal to \textit{state-of-the-art analog accelerators}~\cite{ambrogio_analogai_2023} which perform approximately 12 trillion Operations per second per Watt, equivalent to $\sim 160$~fJ per MAC (or parameter). Even further gains not accessible in digital systems can be realized by biasing our system towards lower-power solutions~\cite{stern_training_2024}. We have demonstrated such gains in the system described in this work~\cite{stern_training_2024}.

\subsection{Learning}
The power-efficiency of our circuit during the learning process is worse than for inference. Each edge's operation of the learning rule as currently configured requires approximately $100$~mW of power, primarily from powering the operational amplifiers (TLV274IN) and analog multiplier (AD633ANZ) integrated circuits, which each require 10s of mW to run. The charging of the gate capacitor is of secondary importance. Here the major component of the power dissipation is across the resistor $R_0 = 100~\Omega$ used by the multiplier circuit to charge the capacitor (as in Twin Edge Circuitry in the appendix). There we have
\begin{equation}
    \frac{\Delta V ^2 }{R_0} = \frac{(V_F^2-V_C^2)^2}{V_0^2 R_0} \leq \frac{V_{max}^4}{V_0^2 R_0}  \sim  4~\text{mW}
\end{equation}
Where $V_{max} = 0.45$~V and $V_0=0.33$~V as defined in the \textit{Twin Edge Circuitry} section of the appendix.

We note that power-efficiency during training is typically less desirable than efficiency during inference~\cite{wright_deep_2022}, and that each contribution to the $100$~mW per edge may be turned off completely during inference.

\section{Companion Circuitry}
\subsection{Clamping Circuitry}

Clamping outputs requires a weighted average of the free network output, and the desired label (Eq.~6:  $O^C = \eta L + (1-\eta) O$), and is shown schematically in Fig.~\ref{fig:workflow}A. For a single output node (non-differential) scheme, a measurement wire is connected to the output node in the free network, measuring $O$. This is fed through a voltage follower (TLV274IN) onto one end of a 10~k$\Omega$ digital potentiometer (AD5220~10k), which is a digitally-adjustable voltage divider. The label $L$ from the DAC is enforced at the other end. The voltage in the middle of this voltage divider is now between $O$ and $L$, creating the nudge. It is fed through another voltage follower (TLV274IN) and onto the output node in the clamped network. Prior to an experiment, the Arduino sets the potentiometer wiper position using $\eta$, allowing the feedback circuit described above to automatically and continuously enforce Eq.~6.

\subsection{Differential Output Labels}
Using a differential output scheme requires an additional step, creating the labels for each node $L_\pm$. This process is shown schematically in Fig~\ref{fig:workflow}B, and uses  multiple amplifiers (TLV274IN) wired as inverting summation amplifiers. These labels are $L_\pm = (O_+ + O_-)/2 \pm L/2$, equivalent to Eq.~7 with $\eta=1$. Each label $L_\pm$ is then fed into its own potentiometer circuit (as in Fig.~\ref{fig:workflow}A), along with the appropriate free network output $O_\pm$, producing the desired clamping voltage $O^C_\pm$, as in Eq.~7. Note that during the experiment, the value of $L_\pm$ is continuously updated using analog feedback from the two output node values $O_+$, $O_-$.

\subsection{Supervisor Circuitry}

\begin{figure*}
\centering
\includegraphics[width=11.4cm]{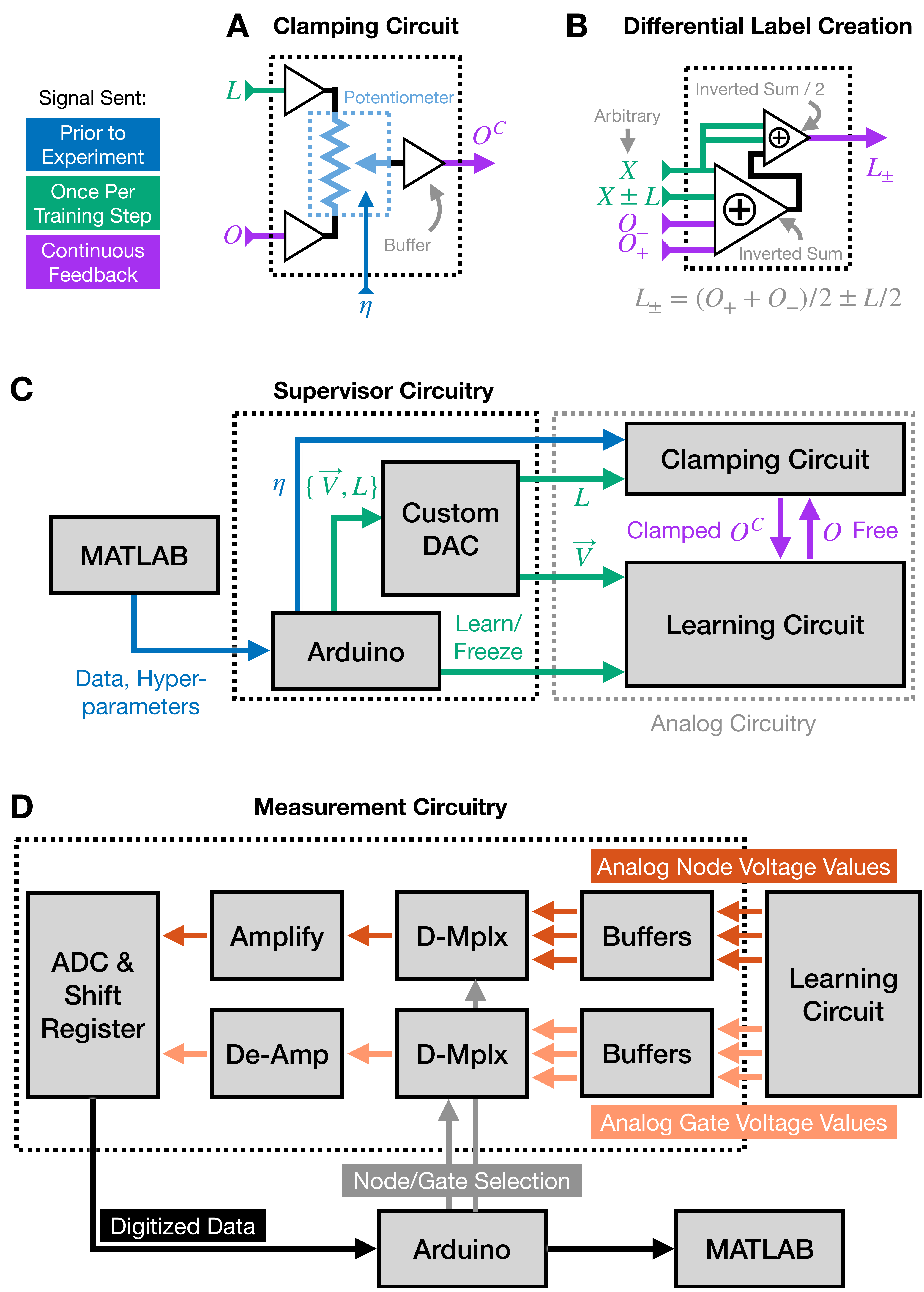}
\caption{\textbf{Companion Circuitry Overview} \textbf{(a) Clamping Circuitry} At the start of the experiment, a digital potentiometer is set using hyperparameter $\eta$. This creates an analog circuit which produces a weighted average of label $L$ (changed every training step) and free output $O$ used to enforce the clamped network output $O_C$. Buffers are voltage followers used to ensure the clamping circuitry does not affect the free network, where $O$ is produced. \textbf{(b) Differential Label Circuitry}. When differential outputs are used, each clamped output $O^C_\pm$ is enforced with its own clamping circuit, as shown in (a) but with $L\rightarrow L_\pm$, $O\rightarrow O_\pm$ and $O^C \rightarrow O^C_\pm$. The labels for each node ($L_+$ and $L_-$) are distinct, and are created using the circuitry shown here. $X$ is an arbitrary value adjusted as needed to keep input voltages to the inverted summing amplifier in the required range ($O<X \pm L<V_{max} = 0.45V$), and does not effect the output. The output of the first inverted summation is $-(2X\pm L + O_+ + O_-)$, and the output of the second yields the formula shown in the figure. \textbf{(c) Supervisor Circuitry} Prior to the experiment, task details are sent to an Arduino Due via serial connection, including hyperparameters and data. Then, the nudge factor $\eta$ (blue) is set, as discussed in (a). For each training step, the supervisor (Arduino) sends inputs $\vec V$ and labels $L$ (green) to a custom DAC (digital to analog converter), which imposes them on the network and clamping circuitry, respectively. Then learning is unfrozen (green) for time $t_h$. Feedback between the learning circuit and the clamping circuit is continuous (purple). \textbf{(d) Measurement Circuitry} When learning is frozen, measurements of the network's internal state are taken in the following manner. All Node and gate voltages are individually buffered and sent into a demultiplexer (D-Mplx). The state of these demultiplexers is chosen by the Arduino. Their output is appropriately amplified or de-amplified to a manageable range for the analog-to-digital converter (ADC), which sends data to the Arduino via a shift register. Note that this measurement process is not required to train or operate the learning circuit (a-c). } 
\label{fig:workflow}
\end{figure*}

The role of supervisor for our system is limited, as the network itself is doing all of the updating. Prior to the experiment, the supervisor sets the nudge hyperparameter $\eta$. Then, at each training step, the only supervisor responsibilities are to apply inputs $\vec V$ to both networks, supply labels $L$ to clamping circuitry, and unfreeze learning for $t_h=100 \mu$s. This cycle is repeated for each datapoint in turn, until a pre-specified number of steps. A schematic of this workflow is shown in Fig.~\ref{fig:workflow}C. 

Because we needed rapid application and measurement of many voltage values, we constructed custom digital-to-analog (DAC) and analog-to-digital (ADC) circuitry, and control the entire process using an Arduino Due.  The clock speed of the Arduino (84~MHz) and the speed with which we can apply individual voltages ($\approx 100$~kHz) are far from state of the art, but much faster than off the shelf-solutions for applying many analog voltages at once. Because the learning step used in this work is $t_h=100 \ \mu s$, slower control hardware would significantly increase the lab-time the system takes to train. As it stands, our control hardware consumes approximately half of the lab-time spent during the learning process (the network evolving is the other half.)

Using custom code on both ends, experiments are designed and parameters (e.g. nudge amplitude $\eta$, train step time $t_h$, total training steps, etc), are chosen in MATLAB, and fed to the Arduino via the USB serial connection. The Arduino then runs an experiment, occasionally passing data back to MATLAB through the serial port, but receiving no further instructions. All applied and measured voltages are 8-bit (0-255), meaning our least significant bit (LSB) for applying voltages to the network is $V_{max}/256 \approx 1.8$~mV. Inputs $V$ and Labels $L$ are passed with a serial digital signal to a shift register (SN74HC595N), then to a digital-to-analog converter (DAC) (TLC7226CN). The DAC converts signals into a 5V range, which are then buffered with an amplifier (TLV274IN) wired to have 1/11 fractional gain, giving us our voltage range for the circuit ($V_{max} \approx 5V/11$). 

While this range is somewhat arbitrary, its size has important effects. The larger this range, the less the relative effect of gate voltages $G$ on conductance compared to average voltage $\overline V$, as the scale of the latter is proportional to the range. That is, a larger range makes the circuit more nonlinear, but also diminishes the importance of the learning response ($G$). The range used allows the learning circuit to meaningfully learn nonlinear tasks, adequately balancing these factors.

Our hardware has the capacity to generate 8 independent analog voltages like the one just described. These 8 can be applied in parallel (but are fed in sequentially), which sets the restriction input nodes + output nodes $\leq 8$. In this work no more than 6 total were used. Inputs are applied directly to the nodes of the network with wires changed by hand for each experiment. Clamped outputs are imposed using the clamping circuitry described above, and are likewise connected to these nodes with wires changed by hand.

Freezing and unfreezing the network is accomplished via a single binary signal that reaches every edge, and is generated using a digital switch (MAX322CPA+). Imposing inputs and labels and freezing/unfreezing the system is the entirety of the supervisor's role in training.

\subsection{Measurement Circuitry}
While not required for operation, we measure the internal state of the network during the training process. At pre-specified test steps, while the network is frozen, and each input set (training or test datapoint) is applied to the network one at a time, and their outputs $O$ measured. The node voltages in both networks and the gate voltages $G$ are also measured. A schematic of the workflow described below is shown in Fig~\ref{fig:workflow}D.

To measure the outputs and the state of the system, we use unity gain buffers (TLV274IN) attached to every node in both networks (16+16) and every gate capacitor (32). These buffers are necessary to prevent our measurement hardware from interfering with the network, as well as to prevent small drains on the capacitors used to store $G$ values that occur with unbuffered measurement. These 64 values are fed into four quadruple 4-to-1 demultiplexers (MAX396CPI+) ($4^3 = 64$), then further demultiplexed 2-to-1 using switches (MAX322CPA+). The Arduino can thus select a single node and a single gate capacitor to measure using a 5-bit signal ($2^5 =32$). The selected node value is buffered and amplified 11x (TLV274IN), whereas the selected gate capacitor is buffered and de-amplified by 2x (TLV274IN). These signals are converted into two 8-bit digital values using an ADC (TLC0820ACN) and fed into two shift registers (CD74HCT597E), which output a serial digital signal read by the Arduino. The least significant bit (LSB) for measuring the nodes is identical to the voltage application LSB ($V_{max}/256 \approx 1.8$~mV), and the LSB for gate voltage $G$ measurement is $10V/256 \approx 40$~mV. Measurements are repeated and averages taken to improve this resolution, but the bound remains close to these values.

\section{Energy-Based Learning Frameworks}
Our system is implemented using the Coupled Learning (CL) framework~\cite{stern_supervised_2021}. CL is closely related to Equilibrium Propagation (EP)~\cite{scellier_equilibrium_2017}, as well as Contrastive Hebbian Learning (CHL)~\cite{movellan_contrastive_1991}. All three implement a contrastive learning scheme in order to train a physical system to perform a task defined by example inputs and outputs. All three involve imposing inputs as boundary conditions, a comparison of a free (inputs only) and a clamped (additional boundary condition at the outputs) phase, and their learning rules are all derived in the same manner from the free and clamped states, up to a constant factor.

The key distinction between the three algorithms is in the boundary conditions imposed on the outputs in the clamped phase. In CHL, the clamped outputs $O^C$ are enforced to be the desired output (labels) $L$. In CL (as in this work), the outputs are enforced to be a weighted average of the free state output values $O^F$ and the desired output $L$. 

\begin{equation}
    O^C_{CHL} = L \quad \quad \quad  O^C_{CL} = (1-\eta)O^F + \eta L 
\end{equation}
Where $0<\eta \leq 1$ is an externally-set parameter, and $\eta=1$ (as we have used for regression tasks in this work) recovers CHL.

In EP, the output (voltage) is not directly enforced, but instead is `nudged' in the direction of the desired output. In an electrical network this amounts to injecting a current $I^C_{EP}$ into the output that is proportional to the difference between the free state output and the desired output. 
\begin{equation}
    I^C_{EP} = \beta(L-O^F)
\end{equation}
Where $0<\beta \ll 1$ is the `nudge parameter.' This formulation in EP allows for analytical guarantees of minimizing the loss function in ideal systems, which CL and CHL lack, however all are empirically successful in many circumstances. For instance, Stern et. al.~\cite{stern_supervised_2021} empirically showed that CL aligns well with gradient descent. Scellier et. al. compares the performance of these three algorithms for several nontrivial computer vision tasks \textit{in silico}~\cite{scellier_energybased_2023a}.

\section{Learning Rule Derivation}
The learning rule of Coupled Learning (and indeed of all three frameworks discussed above) relies on the existence of a quantity $P$ that is minimized by the physical dynamics of the system. The \textit{learning degrees of freedom}, in our case the gate voltage $G$, then evolve proportional to the partial derivative of a difference between $P$ in the free and clamped states:
\begin{equation}
    \dot G \propto -\frac{\partial}{\partial G} (P_C-P_F)
    \label{gdot}
\end{equation}
Linear electronic networks minimize total power dissipation~\cite{vadlamani_physics_2020}, however to generalize to nonlinear networks, a more general quantity must be used. For resistive elements whose current can be written as a function of their voltage drop (\textit{e.g.} resistors and diodes in series), \textit{co-content} (or \textit{pseudo-power} as referred to in~\cite{kendall_training_2020}) may be used: $ P =  \sum_{\text{edges}} \int_0^{\Delta V} I(v) dv$ where $I(v)$ is the current across the element as a function of voltage drop $v$, and $\Delta V$ is the final voltage drop in steady state. Note that for linear resistors $I(v) = vK$ and $P$ evaluates to one half of total power. The derivative of $P$ w.r.t. a given node voltage gives the total current in minus out of a given node, which must sum to zero, and thus $P$ is minimized in steady state. Note that this derivative does not depend on values of $I(v)$ away from equilibrium voltage $v = \Delta V$, and in fact the lower limit of integration $0$ is therefore arbitrary and purely convention.

For our circuit, $I$ cannot be written solely as a function of voltage drop. (Shi, 2003) \cite{shi_pseudoresistive_2003} showed that a network of Ohmic-regime transistors like ours minimize an analytic function (distinct from \textit{pseudo-power}), however only if all transistors have the same gate voltage. However, as our gate voltages vary, and we must make a locally linearizing assumption from which we recover power as the approximate minimization quantity of our circuit. We now quickly sketch this logic using our terminology.

We consider a steady electronic state of one network of transistors as nonlinear variable resistors (\textit{e.g.} the `free' network in our system.) In this state we may write the current through a given edge $i$ (Eq. \ref{conductance} in the main text) as a function of voltage drop $v$ across that edge.
\begin{equation}
I_i(v)  \propto \left[G_i-V_{th}-\overline V_i \right ] v  
\end{equation}
Where $\overline V$ is the average voltage of the edge's two adjacent nodes $\overline V$. This results in a slightly different current function $I_i(v)$ for each edge, but we may now write the locally-valid \textit{pseudo-power} $P^{loc}$, its partial derivative w.r.t. $G_j$, and the derived update rule (Eq. \ref{gdot}) as:

\begin{equation}
\begin{split}
    P^{loc} & = \sum_{\text{i}} \int_0^{\Delta V_i} \underbrace{\Big[G_i-V_{th}-\overline V_i \Big ] v }_{I_i(v)} dv   \\
     \frac{\partial P^{loc}}{\partial G_j} &  = 
     \int_0^{\Delta V_j} v dv =  
      \frac{1}{2}(\Delta V_j)^2 \\
       \dot G_j & \propto (\Delta V_{j,F})^2-(\Delta V_{j,C})^2
    \end{split}
        \label{pseudopower2}
\end{equation}
where we have assumed $\frac{\partial \overline V_i }{\partial G_i} \ll 1$. We find, conveniently, that we recover the same update equation (up to a constant factor) if we naively used total power $P^{tot}$ (as in the main text) instead of our \textit{pseudo-power} derivation:
\begin{equation}
\begin{split}
    \frac{\partial P^{tot}}{\partial G_j} & =  \frac{\partial}{\partial G_j}\sum_{\text{i}} \underbrace{\left[G_i-V_{th}-\overline V_i \right ]\Delta V_i }_{I_i} \cdot  \Delta V_i = (\Delta V_j)^2 \\
    \dot G_j & \propto (\Delta V_{j,F})^2-(\Delta V_{j,C})^2
    \end{split}
\end{equation}
The only sticking point is that to compare the partials of \textit{pseudo-power} in the free and clamped states (Eq. \ref{gdot}) to find useful system updates, the two states must be minimizing the \underline{same} \textit{pseudo-power} equation as each other. That is for each twin edge $i$,
\begin{equation}
\begin{split}
I_{i,F}(v) &\approx I_{i,C}(v)  \\
\left[G_i-V_{th}-\overline V_{i,F} \right ] &\approx \left[G_i-V_{th}-\overline V_{i,C} \right ]
\end{split}
\end{equation}
where we note that since the gate voltage $G_i$ is shared by twin edges it is the same value in both equations. Manipulating this equation we find
\begin{equation}
\begin{split}
1 \approx 1 + \frac{ \overline V_{i,F} - \overline V_{i,C} }{G_i-V_{th}} \\
\left | \overline V_{i,F} - \overline V_{i,C} \right | \ll G_i-V_{th}
\end{split}
\label{approxValid}
\end{equation}
The RHS gives us our region of validity for our learning rule, and nearly always holds in our system for the following reasons. First, gate voltages are often much larger than node voltages. In our system, we connect a voltage $G_{min}=1.1$~V through a diode to our learning rule circuitry such that our gate voltages do not decrease indefinitely. Thus in practice we find our $G$ values always remain above $V_{th}=0.7$~V, and in the vast majority of cases at or above $1$~V, ranging up to $\sim 5.7$~V. This is large compared to the average voltages in the two states $\overline V_F$ and $\overline V_C$, which constrained between $[0,0.45]~V$. Second, the \textit{difference} between these average voltages is smaller still, typically $\left | \overline V_C - \overline V_F \right | \leq 0.1$~V. Third, this difference decreases to zero in the limit of a precisely solved task, or the limit of infinitesimal nudging $\eta \rightarrow 0$. While these limits are not practical, improving a solution and using $\eta < 1$ both increase the range of validity of Eq. \ref{approxValid} and both occur in this work.

We note that in very rare cases (far from solution, large $\eta$, low $G_i$, \underline{and} a large change from clamping on this edge), our learning rule assumptions will not hold. In this case, it is possible (though not guaranteed) that $G$ will evolve in the `incorrect' direction to solve the task. Nevertheless the system consistently finds solutions as we have demonstrated repeatedly in the main text.  Were we to relax our restriction on the lower bounds of $G$, our assumptions might fail more often. However, this change would likely also change the operating characteristics of our transistors (moving from the Triode to Saturation regime), breaking other assumptions of our network's construction as well.

\end{document}